\documentclass[
    ,final            % use final for the camera ready runs
  ]
  {aipproc}

\layoutstyle{6x9}

\begin{document}

\title{ The First Scientific Results from the Pierre Auger Observatory }

\classification{95.85.Ry, 98.70.Sa
%  <Replace this text with PACS numbers; choose from this list:
%                \texttt{http://www.aip..org/pacs/index.html}>
}

\author{T. Yamamoto}
{
  address={KICP, Enrico Fermi Institute, University of Chicago, 5640
  S. Ellis Ave, Chicago IL 60637, USA  }
}

\author{The Pierre Auger Observatory Collaboration}{
  address={}
}

%\author{<author3>}{
%  address={<common address for author2 and author3>}
%  ,altaddress={<author1 address>} % additional visiting address
%}

\begin{abstract}

The southern site of the Pierre Auger Observatory is under the 
construction near Malargue in Argentina and now more than 60\% 
of the detectors are completed. The observatory has been collecting
data for over 1 year and the cumulative exposure is already similar to
that of the largest forerunner experiments.
The hybrid technique provides model-independent energy measurements
from the Fluorescence Detector to calibrate the Surface Detector.
Based on this technique, the first estimation of the energy spectrum
above 3 EeV has been presented and is discussed in this paper.

\end{abstract}

\maketitle

%%%%%%%%%%%%%%%%%%%%%%%%%%%%%%%%%%%%%%%%%%%%
%% MAINMATTER
%%%%%%%%%%%%%%%%%%%%%%%%%%%%%%%%%%%%%%%%%%%%

%\section{Introduction}

The Pierre Auger Observatory is the largest cosmic ray detector
ever built to study the Ultra-High Energy Cosmic Rays
(UHECR) with unprecedented statistics and high precision
\cite{AUGER}. In particular, 
it is important to address whether the cosmic-ray
spectrum continues beyond $10^{20}$ eV. Due to the interaction with microwave
background photons, a steepening is expected around $10^{20}$ eV in the
energy spectrum if the sources are distributed uniformly throughout the
Universe. This conclusion is independent of the composition of the UHECR's.

Recent measurements of the energy spectrum by the AGASA which used
surface detector (SD) array \cite{AGASA} and the HiRes which is using
fluorescence detector (FD) \cite{HyRes} have yielded conflicting
results. There are serious limitations in the use of only the SD or the
FD alone to measure the primary 
spectrum. The SD provides high event statistics with high efficiency
and robust exposure estimation. The SD energy estimation, however,
traditionally relies
on Monte-Carlo simulations which require assumptions about the
hadronic-interaction model and the primary-chemical composition. On the
other hand, the FD provides a calorimetric energy measurement but the
estimation of the exposure has a comparatively large uncertainty
relative to the SD. 

Based on one year operation of a portion of the Pierre Auger
Observatory, the first scientific results were released this
summer concerning the upper limit of the UHE gamma ray flux
\cite{Photon}, anisotropy 
of the arrival directions \cite{Aniso}, and the energy spectrum
\cite{Spect}.  
The cumulative exposure, 1750 $km^2$-$sr$-$yr$, is similar to those
achieved by the largest forerunner experiments. Statistical
uncertainties are still too large to draw any firm conclusions ether
rejecting or confirming results obtained by previous
experiments. However, there is an important step achieved in these results.
The Pierre Auger Observatory was designed as a hybrid detector to observe
the shower particles at ground level by the SD and the associated fluorescence
light generated in the atmosphere by the FD. Combining the strengths of
the SD and the FD, we have developed a reliable estimate of the primary
energy spectrum using the full SD exposure without making assumptions
about the primary masses or hadronic model. \\ 
%In this paper, we discribe the procedure of this spectral analysis.

%\section{Experiment}

The southern site of the Pierre Auger Observatory
is now under construction on an Argentinian pampa ($35^{\circ}$ S,
$69^{\circ}$ W, 1400  m.asl, 875.5 g/cm$^2$). The SD consists of 1600
water Cherenkov tanks planed on a triangular 1.5 km grid covering 3000 $km^2$
area with $2\pi$ sky coverage. The construction of
the Southern site is now 60\% complete. 
%The partial observatory is already the largest
%cosmic ray instrument ever built ($\sim$15 $\times$ AGASA). 
The whole
area of the SD will be overlooked by an FD from 4 sites. Each FD site
has 6 telescopes and each telescope has a $30^{\circ}\times28.6^{\circ}$
field of view with $1.5^{\circ}$ pixel size. Three FD sites are completed
and operating now and one is under construction. 

%\section{Analysis}

The events recorded in the SD are reconstructed using the arrival time
and the signal size from the shower particles reaching the detectors.
The magnitude of the signal at 1 km from the shower axis, S(1000) in
Vertical Equivalent Muon (VEM), is
estimated from the Lateral Distribution Function fit as a size parameter
of the shower \cite{LDF}.
Two cosmic rays of the same energy, but incident at different
zenith angles, will yield different values of S(1000) due to an
attenuation of the shower in the atmosphere. This attenuation is
measured by the well-established technique of the constant intensity cut
(CIC) method. 
The principle of this method is that the nearly isotropic intensity of
cosmic rays means that the integrated intensity above any given energy
must be the same at all zenith angles ($\theta$ degree). One finds the
S(1000) at every zenith angle that corresponds to a single primary
energy by varying S(1000) at each zenith angle to obtain a fixed
integral intensity. 
Based on this method, the zenith angle dependence of S(1000), the CIC
curve is obtained as 
\begin{equation}
S(1000)_{38^{\circ}} = \frac{S(1000)_{\theta}}{1.049+0.0097\theta -
 0.00029\theta^2}
\end{equation}
where $S(1000)_{38^{\circ}}$ VEM is S(1000) adjusted to
$\theta=38^{\circ}$. (The median zenith angle of the showers is
$38^{\circ}$.) 

The link between $S(1000)_{38^{\circ}}$ and the primary energy can be
established using data from the FD. On dark dry nights, the
fluorescence signals are observed simultaneously with the SD 
events. 
%As the FD instruments are used
%primarily as calibration devices in this application, the selection of
%events can be made in a highly selective manner where the FD track had
%to be longer than 350 g cm$^{-2}$, the contribution of the Cherenkov
%light to the signals collected less than 10\% and there were
%contemporaneous measurements of the aerosol content of the
%atmosphere. 
The fit to the FD-energy as a function of $S(1000)_{38^{\circ}}$ is
\begin{equation}
log(E) = -0.79+1.06 log(S(1000)_{38^{\circ}})
\label{eq:energy}
\end{equation}
where $E$ is the FD-energy in EeV. 

\begin{center}
\begin{figure}
  \includegraphics[height=.24\textheight]{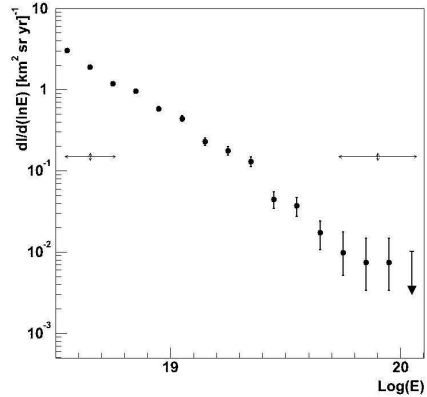}
  \caption{Estimated spectrum. Plotted on the vertical axis is the
 differential flux. Error bars on points indicate statistical
 uncertainty or 95\% CL upper limit. Systematic uncertainty is indicated
 by double arrows at two different energies \cite{Spect}.}
\label{fig:spectrum}
\end{figure}
\end{center}

The events detected by the SD are selected as follows:
The estimated energy must be greater than 3 EeV because detection
efficiency is saturated (nearly 100\%) above this energy. The zenith
angle of the 
arrival direction must be smaller than 60$^{\circ}$. And the event must fall
within a well-defined fiducial area. The estimate of the SD exposure
is simple. The fiducial area is monitored in the trigger system so that
exposure is calculated as the time integration of the aperture given by
the fiducial area and the 60$^{\circ}$ zenith-angle limit.
The spectrum is then obtained by dividing the number of events in given
energy intervals by the exposure as shown in Figure.\ref{fig:spectrum}.

The systematic uncertainty of the energy spectrum comes mainly from the
energy assignment. In the estimation of the FD-energy, there are several
uncertainties which include the fluorescence yield (15\%), missing
energy carried by high-energy muons and neutrinos (4\%), 
the absolute calibration of the FD telescopes (12\%), and
atmospheric condition (10\%). Overall the uncertainty of the FD-energy
is about 25\%. These systematic errors will 
be reduced significantly in a year with completion of the FD
calibration and the measurement of the fluorescence yield in
laboratories . The statistical uncertainty in Equation.\ref{eq:energy}
causes additional energy-dependent systematic uncertainty in the energy
estimation. This uncertainty is dominant to the 
systematic error in the highest energy and will automatically shrink
with the rapidly-increasing hybrid statistics.
The total systematic error is indicated in the
Figure.\ref{fig:spectrum}. 

It should be noted that this energy spectrum 
was measured in the southern sky which could differ from that of
northern sky measured in the previous experiments.
The energy scale based on the FD measurements is systematically lower than
that from an SD analysis that uses QGSJetII simulations with proton
primaries. The difference is similar to the conflicting energy scales of
the HiRes and the AGASA collaborations.
The exposure of the southern observatory is expected to increase by a
factor of 
5$\sim$7 over the next two years. With completion of the FD calibration,
the statistical and systematic errors will shrink accordingly, permitting
a study of spectral features and the energy scale. 

%\begin{theacknowledgments}
This work was supported in part by the Kavli Institute for Cosmological
Physics through the grant NSF PHY-0114422, by NSF AST-0071235, and 
DE-FG0291-ER40606 at the University of Chicago.
%\end{theacknowledgments}

%%%%%%%%%%%%%%%%%%%%%%%%%%%%%%%%%%%%%%%%%%%%%%%%
%% The bibliography can be prepared using the BibTeX program or
%% manually.
%%
%% The code below assumes that BibTeX is used.  If the bibliography is
%% produced without BibTeX comment out the following lines and see the
%% aipguide.pdf for further information.
%%
%% For your convenience a manually coded example is appended
%% after the \end{document}
%%%%%%%%%%%%%%%%%%%%%%%%%%%%%%%%%%%%%%%%%%%%%%%%

%%%%%%%%%%%%%%%%%%%%%%%%%%%%%%%%%%%%%%%%%%%%%%%%
%% You may have to change the BibTeX style below, depending on your
%% setup or preferences.
%%
%%
%% For The AIP proceedings layouts use either
%%%%%%%%%%%%%%%%%%%%%%%%%%%%%%%%%%%%%%%%%%%%

%\bibliographystyle{aipproc}   % if natbib is available

%%%%%%%%%%%%%%%%%%%%%%%%%%%%%%%%%%%%%%%%%%%
%% You probably want to use your own bibtex database here
%%%%%%%%%%%%%%%%%%%%%%%%%%%%%%%%%%%%%%%%%%%
\bibliography{sample}

%%%%%%%%%%%%%%%%%%%%%%%%%%%%%%%%%%%%%%%%%%%
%% Just a reminder that you may have to run bibtex
%% All of it up to \end{document} can be removed
%% if you don't like the warning.
%%%%%%%%%%%%%%%%%%%%%%%%%%%%%%%%%%%%%%%%%%%
\IfFileExists{\jobname.bbl}{}
 {\typeout{}
  \typeout{******************************************}
  \typeout{** Please run "bibtex \jobname" to optain}
  \typeout{** the bibliography and then re-run LaTeX}
  \typeout{** twice to fix the references!}
  \typeout{******************************************}
  \typeout{}
 }

\end{document}